# Transition state redox during dynamical processes in semiconductors and insulators

Guangfu Luo,[1] Thomas F. Kuech,[2] Dane Morgan,[1,*]

[1]Department of Materials Science and Engineering, University of Wisconsin-Madison, Madison, Wisconsin 53706, USA

[2]Department of Chemical and Biological Engineering, University of Wisconsin-Madison, Madison, Wisconsin 53706, USA

*Corresponding author: ddmorgan@wisc.edu

Activation barriers associated with ion diffusion and chemical reactions are vital to understand and predict a wide range of phenomena, such as material growth, ion transport, and catalysis. In the calculation of activation barriers for non-redox processes in semiconductors and insulators, it has been widely assumed that the charge state remains fixed to that of the initial electronic ground state throughout a dynamical process. In this work, we demonstrate that this assumption is generally inaccurate and that a rate-limiting transition state can have a different charge state from the initial ground state. This phenomenon can significantly lower the activation barrier of dynamical process that depends strongly on charge state, such as carbon vacancy diffusion in 4H-SiC. With inclusion of such transition state redox, the activation barrier varies continuously with Fermi level, in contrast to the step-line feature predicted by the traditional fixed-charge assumption. In this study, a straightforward approach to include the transition state redox effect is provided, the typical situations where the effect plays a significant role are identified, and the relevant electron dynamics are discussed.

Keywords: Activation barrier, transition state, redox, semiconductor, insulator, density functional theory

**INTRODUCTION**

Macroscopic dynamical phenomena ranging from ion diffusion to chemical reactions are frequently understood and predicted by analyzing the transition states (TSs) of elementary dynamical processes occurring in extended systems, such as in bulk materials or on bulk material surfaces. A widely-used assumption[1-7] in TS modeling is that the charge state remains fixed during a non-redox process, such as the ion diffusion. Unlike metallic systems, the Fermi level ($E_F$) in the *ab-initio* modeling of an extended semiconducting or insulating system is typically not at the value of the real system, so electron exchange with the bulk states associated with $E_F$ must be included by explicit post-processing. Therefore, the fixed-charge assumption prohibits potential electron exchange between the bulk states and the local region where a dynamical process takes place.

In this article, we remove the fixed-charge assumption and relax the TS charge state to obtain the lowest activation barrier for defect/impurity diffusion in semiconductors and insulators as concrete illustrations. Our density functional theory (DFT) computations confirm that it is energetically favorable for the TS to exchange electrons with the bulk states in several diffusion processes. By allowing such TS redox, the activation barrier is lowered and the $E_F$ dependence of activation barrier becomes continuous. We compare these computational results with available diffusion experiments, analyze the magnitude of the correction associated with the method proposed in this work, and discuss the electron dynamics during the TS redox.

**MODELS AND METHODS**

Our *ab initio* calculations are carried out using DFT as implemented in the Vienna *ab initio* Simulation Package (VASP).[8] An energy cutoff of 400, 450, and 300 eV is set to the plane-wave basis sets for GaAs, 4H-SiC, and Si systems, respectively, and the following projector-augmented wave potentials are utilized: Ga_GW($4s^24p^1$) for Ga, As_GW($4s^24p^3$) for As, Si_GW($3s^23p^2$) for Si, C_GW ($2s^22p^2$) for C, and Li_GW ($2s^1$) for Li. The HSE06[9] hybrid functional is employed, which predicts the band gaps of GaAs, 4H-SiC, and Si to be 1.38, 3.16, 1.15 eV, in good agreement with experimental values[10,11] of 1.42, 3.26, 1.12 eV at 300 K, respectively. The *k*-point sampling is a 3 × 3 × 3 Monkhorst-Pack grid for the GaAs, 4H-SiC, and Si supercells with $a = b = c = 11.20$ Å, $a = b = 9.21$ Å and $c = 10.04$ Å, and $a = b = c = 10.87$ Å, respectively. For defective systems, the defect content is one defect per supercell. The *ab initio* method proposed by Freysoldt, Neugebauer and Van de Walle (FNV)[12] is adopted to remove the image charge interaction and adjust the potential alignment between the perfect and defected structures.

The defect formation energy $E_f$ of a defect $D$ with charge state $q$ is defined as Eqn. 1,[13]

$$E_f(D^q, E_F) \equiv E_{tot}(D^q) + E_{FNV}(D^q) - E_{tot}(bulk) + \sum \mu_i + q[E_{VBM}(bulk) + E_F], \quad (1)$$

where $E_{tot}(D^q)$ and $E_{tot}(bulk)$ are the total internal energies of the systems with defect $D^q$ and perfect bulk, respectively; $E_{FNV}$, $E_{VBM}$ and $E_F$ are the FNV correction, VBM energy and Fermi energy relative to VBM energy, respectively; chemical potentials $\mu_X$ for $X$ = Ga, C, and Li are summarized in the section 4 of the Supplementary Information. The lowest defect formation energy of a defect, $E_f^{min}$, is determined by the minimum value of different charge states as Eqn. 2,

$$E_f^{min}(D, E_F) \equiv \text{Min}\{E_f(D^q, E_F)\}. \quad (2)$$

Defect energy levels correspond to the $E_F$ where the slope of $E_f^{min}$ changes and are independent of the choice of chemical potentials.

The structure and energy of TS is determined using the climbing nudged elastic band (cNEB) method[14]. To relax the TS charge state and obtain the $E_F$ dependence of the lowest activation barrier, a three-step approach is developed here. First, we determine the TS corresponding to a given initial/final charge state by considering the TS charge state within a reasonable range using the cNEB method. Calculations of all possible combinations are demanding. However, from the viewpoint of an energy surface, a transition state does not change when the initial or final state shifts in the same energy valley. For the systems studied here, we indeed find that the TS coordinates and energy are insensitive to the charge state of the initial/final state, so just one initial/final charge state can be used for all the TS charge state calculations if desired (see section 1 of the Supplementary Information). Second, we identify the lowest formation energy curve among these different charges of the TS over the entire range of $E_F$ in the band gap following the same procedure of determining the lowest defect formation energy curve for stable states. Third, we take the difference between the lowest TS energy curve and the initial state energy curve as the $E_F$ dependence of the lowest activation barrier. Two notable consequences of the above relaxed-charge approach are that (1) the activation barrier must be continuous as a function of $E_F$, because the energy curves of TS and initial state are both continuous; (2) the activation barrier must be equal or lower than that obtained from the fixed-charge assumption, because the TS here has the lowest energy in the entire range of $E_F$.

**RESULTS**

We computationally examine three diffusion processes: gallium vacancy, $V_{Ga}$, diffusion in GaAs; carbon vacancy, $V_C$, diffusion in 4H-SiC; and lithium interstitial, Li$_i$, diffusion in silicon. These systems are chosen because they have been extensively studied and thus serve as ideal models to verify our predictions. Additionally, they exemplify three different cases how a TS charge state can differ from that of the initial state, which include electron loss, electron gain, or no electron exchange.

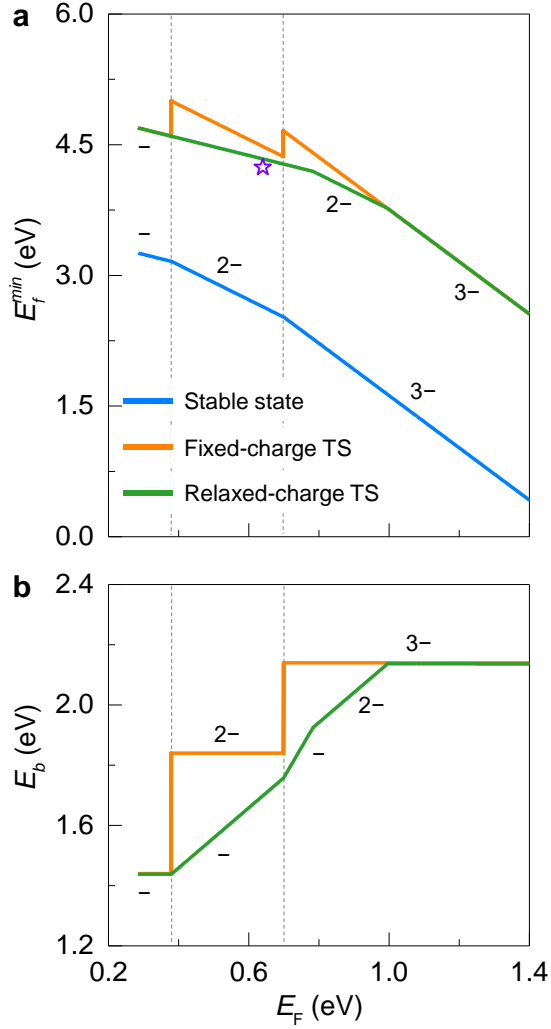

**Figure 1.** Electron loss of TS during the $V_{Ga}$ diffusion in bulk GaAs. (a) Defect formation energy, $E_f^{min}$, of the stable $V_{Ga}$ and its diffusion TS predicted by the fixed-charge and relaxed-charge approaches under As-rich condition. The violet star is an experimental value of the Ga self-diffusion barrier obtained in the temperature range of 800–1225 °C[15]. (b) Comparison of hopping barriers, $E_b$, predicted by the two approaches. $E_F$ is relative to the valence band maximum (VBM). Charge state of each state is labelled. Vertical dashed lines indicate defect energy levels.

**$V_{Ga}$ diffusion in GaAs.** Figure 1a shows the defect formation energy of $V_{Ga}$ and its diffusion TS in GaAs. In the examined range of $E_F$, $V_{Ga}$ consists of three stable charge states from 1− to 3−. Two predicted defect levels at 0.38 and 0.69 eV relative to the VBM are in good agreement with the experimental values[16, 17] of 0.40−0.50 and 0.70−0.75 eV, respectively. Note that $V_{Ga}$ is metastable in the region of $0 \leq E_F \leq 0.28$ eV according to a previous study[18] and the complex diffusion in this range is not investigated here. The TS curve by the relaxed-charge method shows different charge states from those of the stable state in certain regions, indicating that electron exchange of TS with the bulk will occur in these regions. We predict that the electron exchange varies with the $E_F$: specifically, the TS state loses one, then two, and one electron relative to the initial state in the ranges of $0.38 < E_F \leq 0.69$ eV, $0.69 < E_F \leq 0.88$ eV, and $0.88 < E_F \leq 0.90$ eV, respectively. The fixed-charge method is valid only in the heavy n- and p-type doping regions. The hopping barriers in Fig. 1b show that the fixed-charge method predicts a step-line curve with two jumps of 0.39 and 0.31 eV at the boundaries of −/2− and 2−/3−, respectively. In contrast, the relaxed-charge approach significantly lowers the hopping barrier and smoothens the curve in the entire range. The formation energy of TS in Fig. 1a includes the energy to form and migrate the vacancy, and hence corresponds to the Ga collective diffusion barrier, which can be measured in experiments. In the range of $0.38 < E_F \leq 0.90$ eV, the fixed-charge method predicts two prominent spikes, which are reduced by 0 to 0.39 eV with the relaxed-charge method. A compilation of comparison with available diffusion experiments will be discussed in the Discussions section.

**$V_C$ diffusion in 4H-SiC.** Figure 2a shows the defect formation energy of $V_C$ and its diffusion TS in 4H-SiC. Results of the less common silicon vacancy, $V_{Si}$, does not add significant insights and are included in section 2 of the Supplementary Information. Our three predicted defect levels are consistent with previous calculations[19] and the one 0.40 eV below the conduction band minimum (CBM) corresponds to the deep-level-transient-spectroscopy peak $Z_{1/2}$[20-26] observed in the range of 0.59–0.72 eV below CBM, while the two levels 1.30 and 1.58 eV below CBM contribute to the broad peak of $EH_{6/7}$[21, 22, 24, 25, 27-29] in the range of 1.48–1.80 eV below CBM. The TS of $V_C$ can either gain or lose electrons, in contrast to the consistent loss of electron as seen in the TS of $V_{Ga}$. Specifically, the TS gains one, two, and one electron relative to the initial state in the ranges of $0.67 < E_F \leq 1.27$ eV, $1.27 < E_F \leq 1.58$ eV, and $1.58 < E_F \leq 1.86$ eV, respectively; the TS loses two and one electron in the ranges of $2.76 < E_F \leq 2.85$ eV and $2.85 < E_F \leq 3.16$ eV, respectively. Figure 2b shows that the fixed-charge approach induces reductions of hopping barrier in two ranges: 0.67–1.86 eV and 2.76–3.16 eV, and the maximum reduction is up to 1.42 eV at $E_F = 1.58$ eV. As expected, the collective diffusion barrier (formation energy of TS in Fig. 2a) shows significant reduction and is continuous by the relaxed-charge method.

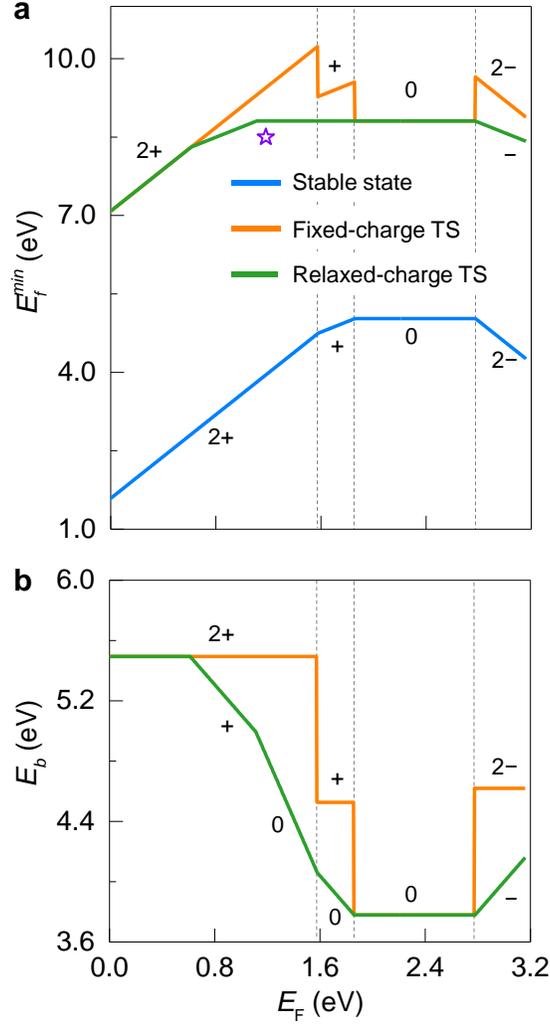

**Figure 2.** Electron gain or loss of TS during the $V_C$ diffusion in bulk 4H-SiC. (a) Defect formation energy, $E_f^{min}$, of the stable $V_C$ and its diffusion TS predicted by the fixed-charge and relaxed-charge approaches under C-rich condition. The violet star is an experimental value of the C self-diffusion barrier obtained in the temperature range of 2100–2350 °C[30]. (b) Comparison of hopping barrier, $E_b$, predicted by the two approaches.

**Li$_i$ diffusion in Si.** Figure 3a shows that Li$_i$ has two stable charge states in bulk Si, + and 0, with a shallow defect level 0.07 eV below the CBM, which is in excellent agreement with the experimental value of 0.03 eV below the CBM[31]. In contrast to the $V_{Ga}$ in GaAs and $V_C$ in 4H-SiC, Li$_i$ and its TS possess almost the same charge states (Fig. 3a) and therefore the fixed- and relaxed-charge approaches yield almost the same activation barriers. Figure 3b shows that the hopping barriers of Li$_i^+$ and Li$_i^0$ are predicted to be 0.632 and 0.625 eV, respectively.

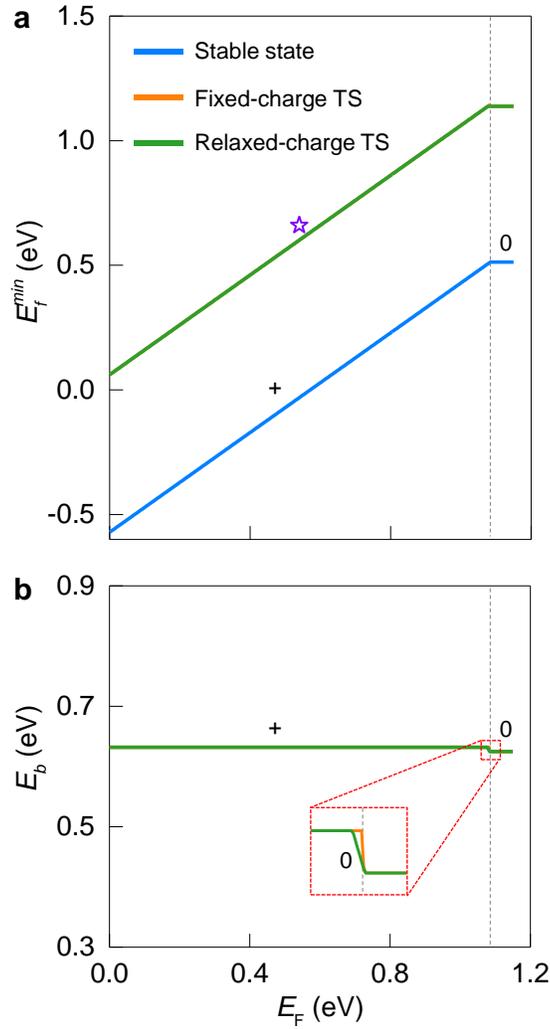

**Figure 3.** Electron gain of TS during the Li$_i$ diffusion in bulk Si. (a) Defect formation energy, $E_f^{min}$, of the stable Li$_i$ and its diffusion TS predicted by the fixed-charge and relaxed-charge approaches. Bulk Li is used the chemical potential of Li. Note that the orange and green curves are almost identical. The violet star is an experimental value of the Li$_i$ diffusion barrier obtained in the temperature range of 25–125 ˚C[32, 33]. (b) Comparison of hopping barrier, $E_b$, predicted by the two approaches.

## DISCUSSIONS

**Comparison with diffusion experiments.** Figure 1a and 2a indicate previous experimental values of Ga and C self-diffusion barriers through the vacancy-mediated mechanism in GaAs and 4H-SiC, respectively. The diffusion barrier in intrinsic GaAs was measured to be 4.24 eV under As-rich condition in the temperature range of 800–1225 ˚C[15] and the diffusion barrier in intrinsic 4H-SiC was 8.50 eV under C-rich condition in the temperature range of 2100–2350 ˚C[30] (see section 3 of the Supplementary Information for determination of the $E_F$ in experiments). In both cases the relaxed-charge approach agrees with the experiments better than the fixed-charge approach, which generally supports the former. However, robust experimental verification of the improvements by the relaxed-charge method is not straightforward, because the high experimental temperatures induce significantly reduced band gaps, substantial thermal vibrational energies, and increasing contribution of excited states, all of which could be important to the correct prediction of activation barrier. For instance, relative to the values at 300 K, the temperatures of 800–1225 °C reduce the band gap of GaAs by 0.39–0.61 eV, and the temperatures of 2100–2350 °C reduce the band gap of 4H-SiC by 0.91–1.04 eV. Therefore, future experiments at relatively lower temperatures or theoretical thermal corrections to the current results would be valuable to enable compelling validation of the predictions.

Unlike the two other cases, low-temperature experiments for $Li_i$ diffusion in intrinsic Si[32, 33] had been performed. The experimental temperatures of 25–125 °C correspond to Si band gaps of 1.13–1.10 eV, which are very close to the value of 1.15 eV in our simulation. Our predicted collective diffusion barrier of 0.60 eV at $E_F = 0.54$ eV, which corresponds to 75 °C for intrinsic Si (section 3 of the Supplementary Information), is in good agreement with the experimental value of 0.66 eV, as shown in Fig. 3a. Unfortunately, the fixed- and relaxed-charge methods yield almost the same predictions in this case, so this comparison cannot be used to distinguish the two approaches.

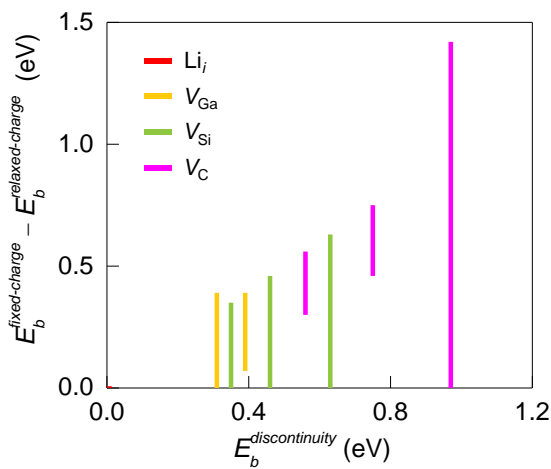

**Figure 4.** Difference of hopping barrier between the fixed-charge and relaxed-charge methods versus the hopping barrier discontinuity predicted by the fixed-charge method.

**Magnitude of correction by the relaxed-charge method.** To gain insights into when the TS redox is likely to play a significant (minor) role, we compare the correction of hopping barrier by the relaxed-charge method with the discontinuity of hopping barrier predicted by the fixed-charge method. Figure 4 shows that the maximum correction of hopping barrier generally increases with the discontinuity. In other words, the relaxed-charge method plays a significant (minor) role when the fixed-charged method predicts a strong (weak) charge dependence. Physically, this correlation arises from the fact that a discontinuity is caused by forcing the TS into an incorrect charge state, so a larger discontinuity corresponds to a larger error in the TS energy and more reduction when this error is corrected.

It is unclear what combination of migrating species and host material would exhibit strong charge dependence, although it is known that a charge state can impact activation energy through the change of ionic size and bonding characteristics. A previous study[34] has found that the charge sensitivity of activation barrier depends linearly on the defect level shift in the TS relative to that in the initial state, a relationship can be easily understood with Fig. S3 in the Supplementary Information. However, such defect level shift requires essentially the same calculations as a traditional activation barrier and thus does not enable easy prediction.

**Electron dynamics during TS redox.** Because all our results so far are based on thermodynamic equilibrium, it is interesting to investigate how the electron dynamics could influence the TS redox. Taking the hopping with an oxidized TS as an example, Fig. 5 schematically shows the time scales of four relevant processes, together with schematic suggestions of where the processes take place: (1) the time residing in initial states before a successful hop, (2) the ionic hop time, (3) the time for electrons to leave the ion or the oxidation time, and (4) the time for the electrons to combine with the ion or the reduction time. Given that most dynamical processes possess activation barriers greater than 0.3 eV and most experimental temperatures are less than 3000 ºC, a reasonable lower limit of time scale 1 is many picoseconds, according to the Arrhenius equation and an attempt frequency of $5\times10^{12}$ Hz. The time scale 2 is generally on the scale of 0.1 picosecond[35]. The time scales 3 and 4 have not been established and may depend strongly on the hopping species and the host material. For example, the reduction process could be similar to electron-hole recombination, which can occur in the wide time scale from nanoseconds to microsceconds[36-41].

Depending on the four time scales, the redox during the ionic hop in Fig. 5 could involve two possible mechanisms. Since the different charge states in the initial state are largely in thermal equilibrium (time scale 1 >> time scale 2), one mechanism is that a thermally excited ion with charge state $q_2$ hops with constant charge. This pathway (orange curve) is dominant because it possesses the lowest collective diffusion barrier of $E_{TS}^0 - E_{initial}^0$, where $E_{TS}^0$ and $E_{initial}^0$ are the energies of the rate-limiting TS and initial ground state, respectively. If the redox are much faster than the ionic hop (time scales 3 and 4 << time scale 2), an additional mechanism that involves the variation from initial state $q_1$ to the TS $q_2$ and back to final state $q_1$ will be invoked, which can be realized through electron emission/absorption during the ionic hop. Both mechanisms predict the same collective diffusion barrier as that based solely on thermodynamic equilibrium. Note that the electron emission/absorption during a hop is not equivalent to the charge redistribution caused by rehybridization in a traditional TS calculation. The rehybridization may play a significant role in determining the activation energy, such as for oxygen vacancy hopping in some perovskite oxides,[42] but it does not involve electron emission/absorption and does not require the special treatments described in this work.

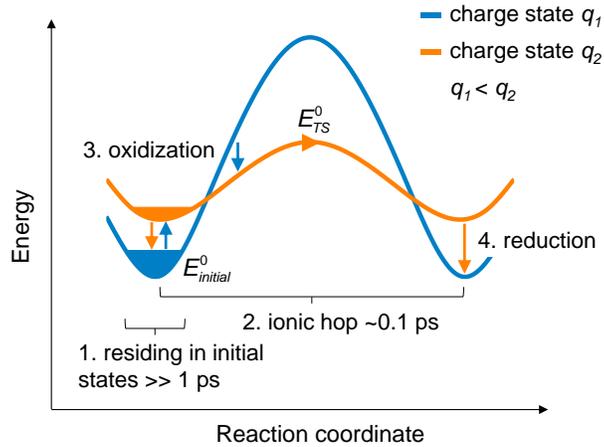

**Figure 5.** Schematic of four processes involved in a migration where a rate-limiting TS loses electrons relative to the initial ground state.

**SUMMARY**

In summary, we have developed a relaxed-charge approach to study activation barriers in semiconductors and insulators, and identified several diffusion processes, in which a TS redox can lower the activation barrier. The TS redox results in continuous dependence of activation barrier with $E_F$, in contrast to the step-line feature predicted by the traditional fixed-charge approach. This effect is particularly important to dynamical processes where the activation barriers are sensitive to charge state. The TS redox can be realized through two possible mechanisms: the constant-charge hopping of a thermally excited initial state and the varying-charge hopping of an initial ground state. It should be noted that the computational procedure proposed in this work is also applicable to processes that have different charge states between the initial and final states. In view of the fundamental role of activation barriers, the possibly significant overestimation by the fixed-charge method, and the straightforward new computational approach, we suggest including the possibility of TS redox in future prediction of dynamical processes in semiconductors and insulators.

**ACKNOWLEDGMENTS**

This research was primarily supported by NSF through the University of Wisconsin Materials Research Science and Engineering Center (Grant No. DMR-1121288). Computations in this work benefited from the use of the Extreme Science and Engineering Discovery Environment (XSEDE), which is supported by National Science Foundation Grant No. ACI-1053575, the computing resources and assistance of the UW-Madison Center For High Throughput Computing (CHTC) in the Department of Computer Sciences, and the National Energy Research Scientific Computing Center (NERSC), a DOE Office of Science User Facility supported by the Office of Science of the U.S. Department of Energy under Contract No. DE-AC02-05CH11231.

**CONFLICT OF INTEREST**

The authors declare no conflict of interest.

**References**

1. Lee, W. C., Lee, S. G. & Chang, K. J. First-principles study of the self-interstitial diffusion mechanism in silicon. *J. Phys.-Condes. Matter* **10**, 995-1002 (1998).
2. Eichler, A. Tetragonal Y-doped zirconia: structure and ion conductivity. *Phys. Rev. B* **64**, 174103 (2001).
3. Seebauer, E. G. & Kratzer, M. C. *Charged semiconductor defects: structure, thermodynamics and diffusion*. Springer (2009).
4. Eames, C., Frost, J. M., Barnes, P. R. F., O'Regan, B. C., Walsh, A. & Islam, M. S. Ionic transport in hybrid lead iodide perovskite solar cells. *Nat. Commun.* **6**, 7497 (2015).
5. Tao, X. Y.*, et al.* Balancing surface adsorption and diffusion of lithium-polysulfides on nonconductive oxides for lithium-sulfur battery design. *Nat. Commun.* **7**, 11203 (2016).
6. Gray, C., Lei, Y. K. & Wang, G. F. Charged vacancy diffusion in chromium oxide crystal: DFT and DFT plus U predictions. *J. Appl. Phys.* **120**, 215101 (2016).
7. Medasani, B., Sushko, M. L., Rosso, K. M., Schreiber, D. K. & Bruemmer, S. M. Vacancies and vacancy-mediated self diffusion in $Cr_2O_3$: a first-principles study. *J. Phys. Chem. C* **121**, 1817-1831 (2017).
8. Kresse, G. & Furthmuller, J. Efficient iterative schemes for *ab initio* total-energy calculations using a plane-wave basis set. *Phys. Rev. B* **54**, 11169-11186 (1996).
9. Heyd, J., Scuseria, G. E. & Ernzerhof, M. Hybrid functionals based on a screened Coulomb potential. *J. Chem. Phys.* **118**, 8207-8215 (2003).
10. Sze, S. M. & Ng, K. K. *Physics of semiconductor devices*, 3rd edn. Wiley-Interscience (2007).
11. Choyke, W. J., Hamilton, D. R. & Patrick, L. Optical properties of cubic SiC: luminescence of nitrogen-exciton complexes, and interband absorption. *Phys. Rev.* **133**, A1163 (1964).
12. Freysoldt, C., Neugebauer, J. & Van de Walle, C. G. Fully *ab initio* finite-size corrections for charged-defect supercell calculations. *Phys. Rev. Lett.* **102**, 016402 (2009).
13. Freysoldt, C.*, et al.* First-principles calculations for point defects in solids. *Rev. Mod. Phys.* **86**, 253 (2014).
14. Henkelman, G., Uberuaga, B. P. & Jonsson, H. A climbing image nudged elastic band method for finding saddle points and minimum energy paths. *J. Chem. Phys.* **113**, 9901-9904 (2000).
15. Wang, L.*, et al.* Ga self-diffusion in GaAs isotope heterostructures. *Phys. Rev. Lett.* **76**, 2342 (1996).

16. Milnes, A. G. Impurity and defect levels (experimental) in gallium arsenide. *Adv. Electron. Electron Phys.* **61**, 63-160 (1983).
17. Gebauer, J., *et al.* Determination of the Gibbs free energy of formation of Ga vacancies in GaAs by positron annihilation. *Phys. Rev. B* **67**, 235207 (2003).
18. Luo, G. F., Yang, S. J., Jenness, G. R., Song, Z. W., Kuech, T. F. & Morgan, D. Understanding and reducing deleterious defects in the metastable alloy GaAsBi. *NPG Asia Mater.* **9**, e345 (2017).
19. Weber, J. R., *et al.* Quantum computing with defects. *Proc. Natl. Acad. Sci. U. S. A.* **107**, 8513-8518 (2010).
20. Dalibor, T., *et al.* Deep defect centers in silicon carbide monitored with deep level transient spectroscopy. *Phys. Status Solidi A* **162**, 199-225 (1997).
21. Storasta, L., Bergman, J. P., Janzen, E., Henry, A. & Lu, J. Deep levels created by low energy electron irradiation in 4H-SiC. *J. Appl. Phys.* **96**, 4909-4915 (2004).
22. Alfieri, G., Monakhov, E. V., Svensson, B. G. & Linnarsson, M. K. Annealing behavior between room temperature and 2000 $^{o}$C of deep level defects in electron-irradiated *n*-type 4H silicon carbide. *J. Appl. Phys.* **98**, 043518 (2005).
23. Castaldini, A., Cavallini, A. & Rigutti, L. Assessment of the intrinsic nature of deep level $Z_1/Z_2$ by compensation effects in proton-irradiated 4H-SiC. *Semicond. Sci. Tech.* **21**, 724-728 (2006).
24. Danno, K. & Kimoto, T. Investigation of deep levels in n-type 4H-SiC epilayers irradiated with low-energy electrons. *J. Appl. Phys.* **100**, 113728 (2006).
25. Son, N. T., *et al.* Negative-U system of carbon vacancy in 4H-SiC. *Phys. Rev. Lett.* **109**, 187603 (2012).
26. Kawahara, K., Trinh, X. T., Son, N. T., Janzen, E., Suda, J. & Kimoto, T. Quantitative comparison between $Z_{1/2}$ center and carbon vacancy in 4H-SiC. *J. Appl. Phys.* **115**, 143705 (2014).
27. Hemmingsson, C., *et al.* Deep level defects in electron-irradiated 4H SiC epitaxial layers. *J. Appl. Phys.* **81**, 6155-6159 (1997).
28. Son, N. T., Magnusson, B. & Janzen, E. Photoexcitation-electron-paramagnetic-resonance studies of the carbon vacancy in 4H-SiC. *Appl. Phys. Lett.* **81**, 3945-3947 (2002).
29. Booker, I. D., Janzen, E., Son, N. T., Hassan, J., Stenberg, P. & Sveinbjornsson, E. O. Donor and double-donor transitions of the carbon vacancy related $EH_{6/7}$ deep level in 4H-SiC. *J. Appl. Phys.* **119**, 235703 (2016).
30. Linnarsson, M. K., Janson, M. S., Zhang, J., Janzen, E. & Svensson, B. G. Self-diffusion of $^{12}$C and $^{13}$C in intrinsic 4H-SiC. *J. Appl. Phys.* **95**, 8469-8471 (2004).
31. Chen, J. W. & Milnes, A. G. Energy levels in silicon. *Annu. Rev. Mater. Sci.* **10**, 157-228 (1980).
32. Pell, E. M. Diffusion rate of Li in Si at low temperatures. *Phys. Rev.* **119**, 1222 (1960).
33. Canham, L. T. *Properties of silicon*. INSPEC, Institution of Electrical Engineers (1988), pp 455-457.
34. Lei, Y. K. & Wang, G. F. Linking diffusion kinetics to defect electronic structure in metal oxides: Charge-dependent vacancy diffusion in alumina. *Scripta Mater.* **101**, 20-23 (2015).
35. Zewail, A. H., Schryver, F. C. D., Feyter, S. D. & Schweitzer, G. *Femtochemistry: with the Nobel lecture of A. Zewail*. Wiley-VCH (2001).
36. Schmid, W. Auger lifetimes for excitons bound to neutral donors and acceptors in Si. *Phys. Status Solidi B* **84**, 529-540 (1977).
37. Thooft, G. W., Vanderpoel, W. A. J. A., Molenkamp, L. W. & Foxon, C. T. Giant oscillator strength of free excitons in GaAs. *Phys. Rev. B* **35**, 8281-8284 (1987).
38. Cancio, A. C. & Chang, Y. C. Quantum Monte Carlo studies of binding energy and radiative lifetime of bound excitons in direct-gap semiconductors. *Phys. Rev. B* **47**, 13246-13259 (1993).
39. Egilsson, T., Bergman, J. P., Ivanov, I. G., Henry, A. & Janzen, E. Properties of the $D_1$ bound exciton in 4H-SiC. *Phys. Rev. B* **59**, 1956-1963 (1999).

40. Hain, T. C., *et al.* Excitation and recombination dynamics of vacancy-related spin centers in silicon carbide. *J. Appl. Phys.* **115**, 133508 (2014).
41. Bracher, D. O., Zhang, X. Y. & Hu, E. L. Selective purcell enhancement of two closely linked zero-phonon transitions of a silicon carbide color center. *Proc. Natl. Acad. Sci. U. S. A.* **114**, 4060-4065 (2017).
42. Mastrikov, Y. A., Merkle, R., Kotomin, E. A., Kuklja, M. M. & Maier, J. Formation and migration of oxygen vacancies in $La_{1-x}Sr_xCo_{1-y}Fe_yO_{3-\delta}$ perovskites: insight from ab initio calculations and comparison with $Ba_{1-x}Sr_xCo_{1-y}Fe_yO_{3-\delta}$. *Phys. Chem. Chem. Phys.* **15**, 911 (2013).

**FIGURE CAPTIONS**

**Figure 1.** Electron loss of TS during the $V_{Ga}$ diffusion in bulk GaAs. (a) Defect formation energy, $E_f^{min}$, of the stable $V_{Ga}$ and its diffusion TS predicted by the fixed-charge and relaxed-charge approaches under As-rich condition. The violet star is an experimental value of the Ga self-diffusion barrier obtained in the temperature range of 800–1225 °C[15]. (b) Comparison of hopping barriers, $E_b$, predicted by the two approaches. $E_F$ is relative to the valence band maximum (VBM). Charge state of each state is labelled. Vertical dashed lines indicate defect energy levels.

**Figure 2.** Electron gain or loss of TS during the $V_C$ diffusion in bulk 4H-SiC. (a) Defect formation energy, $E_f^{min}$, of the stable $V_C$ and its diffusion TS predicted by the fixed-charge and relaxed-charge approaches under C-rich condition. The violet star is an experimental value of the C self-diffusion barrier obtained in the temperature range of 2100–2350 °C[30]. (b) Comparison of hopping barrier, $E_b$, predicted by the two approaches.

**Figure 3.** Electron gain of TS during the $Li_i$ diffusion in bulk Si. (a) Defect formation energy, $E_f^{min}$, of the stable $Li_i$ and its diffusion TS predicted by the fixed-charge and relaxed-charge approaches. Bulk Li is used the chemical potential of Li. Note that the orange and green curves are almost identical. The violet star is an experimental value of the $Li_i$ diffusion barrier obtained in the temperature range of 25–125 °C[32, 33]. (b) Comparison of hopping barrier, $E_b$, predicted by the two approaches.

**Figure 4.** Difference of hopping barrier between the fixed-charge and relaxed-charge methods versus the hopping barrier discontinuity predicted by the fixed-charge method.

**Figure 5.** Schematic of four processes involved in a migration where a rate-limiting TS loses electrons relative to the initial ground state.

# Supplemental Information for "Transition state redox during dynamical processes in semiconductors and insulators"

Guangfu Luo, Thomas F. Kuech, Dane Morgan

## 1. Insensitivity of transition states to the charge states of initial and final states

To identify the transition state (TS) in the entire range of $E_F$, we systematically examined several possible TS charge states for each initial and final state. These calculations are significantly heavier than the fixed-charge method. For example, for a defect with $m$ stable charge states and considering $n$ possible TS charge states for each initial charge state, one needs to carry out totally $m \times n$ calculations, in contrast to the $m$ calculations of the fixed-charge method. Here, we show that the TS is insensitive to the charge states of the initial and final states for at least our examined defects ($V_{Ga}$ in GaAs, $V_C$ and $V_{Si}$ in 4H-SiC, and $Li_i$ in Si). As shown in Table S1, TSs obtained from different charge states of initial states differ by energies less than 8 meV. As a result, the total number of calculations can be reduced from $m \times n$ to roughly $m+n$ by doing only one calculation for each TS charge state.

**Table S1.** Maximum energy differences, $\delta$ (meV), among TSs calculated using different initial states. $E_F$ is set to valance band maximum; $q_{TS}$ is the charge state of TS and $q_{initial}$ is the examined range of charge sates of initial states.

| $V_{Ga}$ hopping in GaAs | | $V_C$ hopping in 4H-SiC | | $V_{Si}$ hopping in 4H-SiC | | $Li_i$ hopping in Si | |
|---|---|---|---|---|---|---|---|
| ($q_{TS}$, $q_{initial}$) | $\delta$ | ($q_{TS}$, $q_{initial}$) | $\delta$ | ($q_{TS}$, $q_{initial}$) | $\delta$ | ($q_{TS}$, $q_{initial}$) | $\delta$ |
| (-3, [-3, -2]) | 4 | (-2, [-2, -1]) | 8 | (-3, [-3, -2]) | 7 | (1, [0, 1]) | 0 |
| (-2, [-3, -1]) | 1 | (-1, [-2, 0]) | 8 | (-2, [-3, -1]) | 3 | (0, [-1, 1]) | 0 |
| (-1, [-3, -1]) | 8 | (0, [-1, 2]) | 4 | (-1, [-2, 0]) | 1 | (-1, [-1, 0]) | 0 |
| (0, [-3, -1]) | 8 | (1, [0, 2]) | 7 | (0, [-2, 0]) | 0 | | |
| | | (2, [1, 2]) | 1 | | | | |

This insensitivity of the TSs arises from the fact that the geometrical structures of the initial and final states are largely insensitive to the charge states. Figure S1 shows that for all the examined initial states, the largest atomic displacement induced by different charge states is less than 0.09 Å. These small structural differences also lead to relatively small energy differences. For example, the total energy of $V_{Ga}^-$ ($V_{Ga}^{2-}$) using the relaxed geometrical structure of $V_{Ga}^{2-}$ ($V_{Ga}^-$) is only 70 (90) meV higher than that obtained from the relaxed structure of $V_{Ga}^-$ ($V_{Ga}^{2-}$). Therefore, a defect with different charge states are in the same energy basin, and different initial/final states are expected to lead to the same TS, at least in theory.

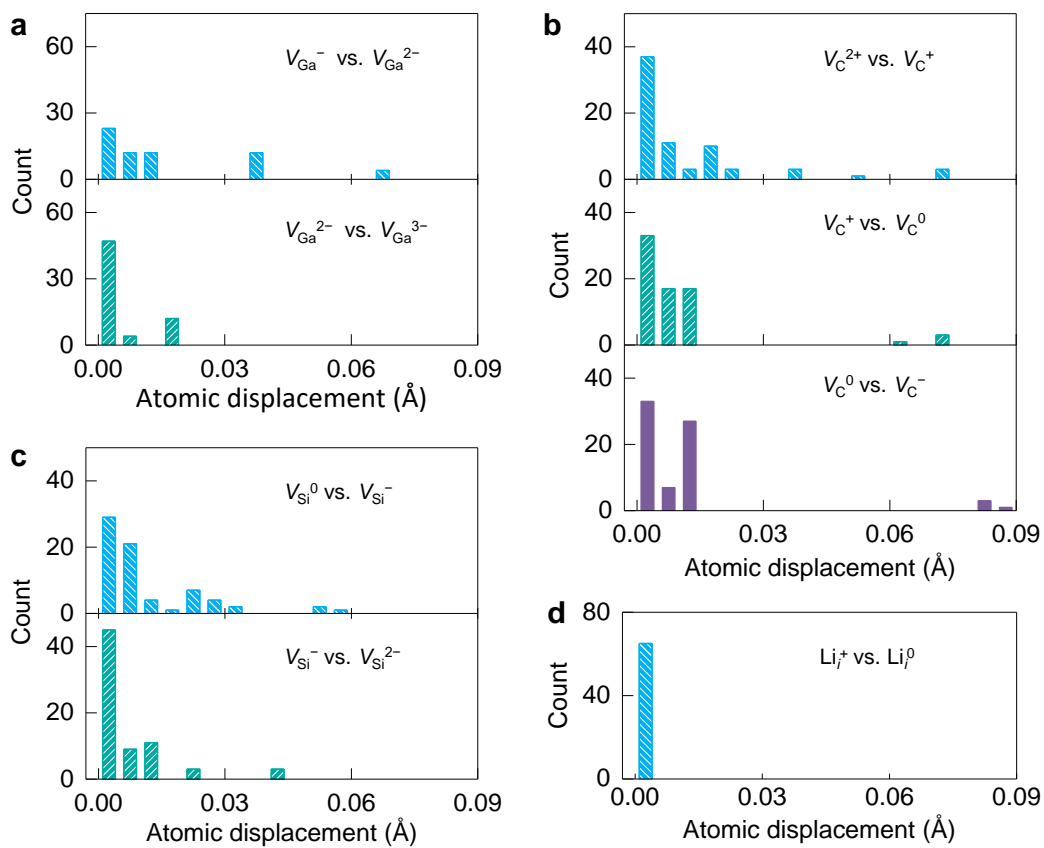

**Figure S1.** Histogram of atomic displacements between different charge states for (a) $V_{Ga}$ in bulk GaAs, (b) $V_C$ and (c) $V_{Si}$ in bulk 4H-SiC, and (d) Li$_i$ in bulk Si.

S2

## 2. Results of $V_{Si}$ in 4H-SiC

$V_{Si}$ is predicted to possess three defect levels at 1.28, 2.39, and 2.89 eV above the valance band maximum (Fig. S2a). The hopping barriers in Fig. S2b show that the fixed-charge method predicts a step-line curve with three jumps of 0.63, 0.46, and 0.35 eV at the boundaries of 0/−, −/2− and 2−/3−, respectively. By contrast, the relaxed-charge approach lowers the hopping barrier and smoothens the curve in the entire region. We predict that the TS gains one electron relative to the initial state in three ranges: $0.65 < E_F \leq 1.28$ eV, $1.93 < E_F \leq 2.39$ eV, and $2.54 < E_F \leq 2.89$ eV.

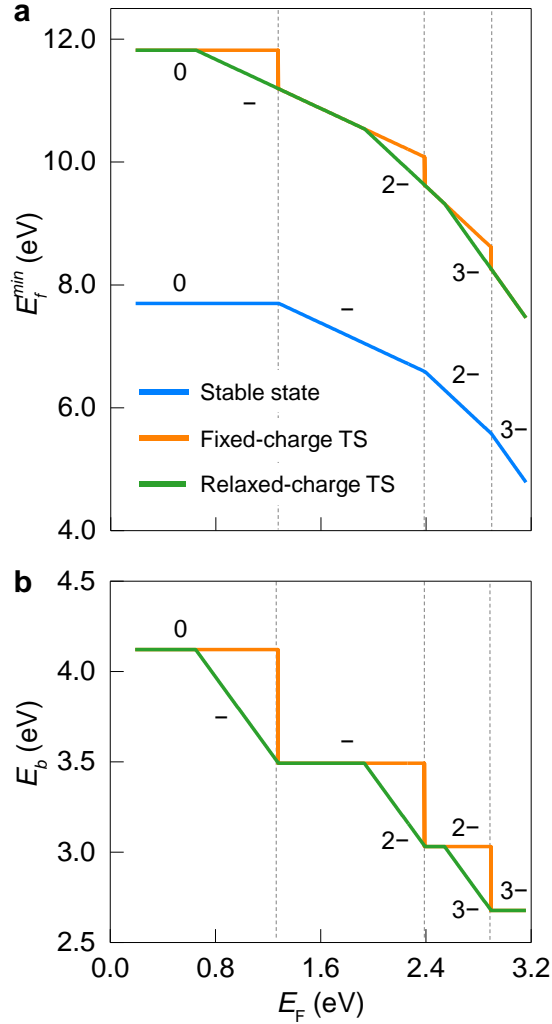

**Figure S2.** Electron gain of TS during the $V_{Si}$ diffusion in bulk 4H-SiC. (a) Defect formation energy, $E_f^{min}$, of the stable $V_{Si}$ and its diffusion TS predicted by the fixed-charge and relaxed-charge approaches under C-rich condition. (b) Comparison of hopping barrier, $E_b$, predicted by the two approaches. $E_F$ is relative to the valence band maximum. Charge state of each state is labelled. Vertical dashed lines indicate defect energy levels.

## 3. Estimation of $E_F$ for intrinsic GaAs, 4H-SiC, and Si

Based on the well-known relationship of Eqn. S1[1], the effective density-of-state masses, and the band gap relationship with temperature, the calculated $E_F$ at particular temperatures for intrinsic GaAs, 4H-SiC, and Si are listed in Table S2.

$$E_F(T) = \frac{1}{2}\left(E_g(T) + k_B T \ln(\frac{m_h^{*3/2}}{m_e^{*3/2}})\right) \quad (S1)$$

**Table S2.** Effective density-of-state mass for electron ($m^*_e$) and hole ($m^*_h$), temperature dependence of band gap $E_g(T)$, and calculated band gap and $E_F$ at certain temperatures for intrinsic GaAs, 4H-SiC, and Si. Units of effective mass and energy are the rest mass of electron and eV, respectively.

|        | ($m^*_e$, $m^*_h$) | $E_g(T)$, $T$ in K | ($E_g$, $E_F$) at $T_0$ |
|--------|--------------------|--------------------|-----------------|
| GaAs   | (0.06, 0.52) †     | $1.52 - 5.40\times10^{-4} T^2/(T+204)$ † | (0.93, 0.64) at 1000 °C |
| 4H-SiC | (0.77, 1.00) ‖     | $3.27 - 6.00\times10^{-4} T^2/(T+1200)$ ‖ | (2.27, 1.18) at 2200 °C |
| Si     | (1.08, 0.55) †     | $1.17 - 4.90\times10^{-4} T^2/(T+655)$ † | (1.11, 0.54) at 75 °C |

† Ref. 2
‖ Ref. 3

## 4. Chemical potentials for calculations of defects in GaAs, 4H-SiC, and Si

**Table S3.** Chemical potentials $\mu$ (eV) of Ga, C and Li under As-rich, C-rich, and Li-rich conditions for the energy calculations of $V_{Ga}$ in GaAs, $V_C$ and $V_{Si}$ in 4H-SiC, and $Li_i$ in Si. $\mu_{Ga}$ = $E_{tot}$(per formula of GaAs bulk) – $E_{tot}$(per atom of hexagonal As bulk); $\mu_C$ = $E_{tot}$(per atom of graphite); $\mu_{Si}$ = $E_{tot}$(per formula of 4H-SiC bulk) – $E_{tot}$(per atom of graphite); $\mu_{Li}$ = $E_{tot}$(per atom of body-centered cubic Li bulk).

| $\mu_{Ga}$ under As-rich | $\mu_C$ under C-rich | $\mu_{Si}$ under C-rich | $\mu_{Li}$ under Li-rich |
|---|---|---|---|
| −4.58 | -10.61 | -6.82 | -1.97 |

## 5. Linear dependence of activation barrier change with defect level shift

As shown in Fig. S3, the activation barrier change $\delta E_b$ in two charge states depends linearly on the shift of defect level $\delta E_l$ through Eqn. S2.

$$\delta E_b = \delta E_l (\tan\theta^{q2} - \tan\theta^{q1}) \tag{S2}$$

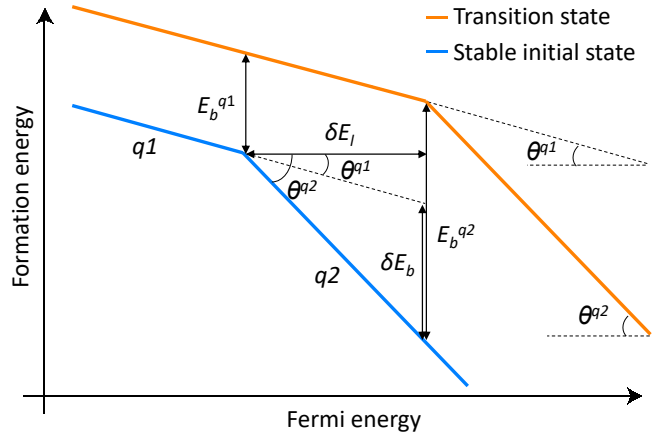

**Figure S3.** Relationship between activation barrier change, $\delta E_b$, and defect level shift, $\delta E_l$, in two charge states. The activation barrier of charge state $q1$ ($q2$) is $E_b^{q1}$ ($E_b^{q2}$) and the angle between the segment $q1$ ($q2$) relative to the horizontal axis is $\theta^{q1}$ ($\theta^{q2}$).

## References

1. Pierret, R. F. *Advanced semiconductor fundamentals*, 2nd edn. Prentice Hall (2003).
2. Sze, S. M. & Ng, K. K. *Physics of semiconductor devices*, 3rd edn. Wiley-Interscience (2007).
3. http://www.ioffe.ru/SVA/NSM/Semicond/index.html.